\documentclass[12pt]{iopart}

 \usepackage{graphicx}
 \usepackage{wrapfig}
\begin{document}

\title[The first measurement of $B$ contribution at RHIC]{The first measurement of $B$ meson semi-leptonic decay contribution to non-photonic electrons at RHIC}

\author{Xiaoyan Lin ({\it for the STAR collaboration})}

\address{Institute of Particle Physics, Central China Normal University, Wuhan 430079, China }
\ead{linxy@iopp.ccnu.edu.cn}
\begin{abstract}
We present the first measurement for the $B$ meson semi-leptonic
decay contribution to non-photonic electrons at RHIC using
non-photonic electron azimuthal correlations with charged hadrons in
p+p collisions at $\sqrt{s_{NN}} = 200$ GeV from STAR.
\end{abstract}


\section{Introduction}
\vspace{-0.3cm} Recent experimental data~\cite{star_e,phenix_e} from
RHIC show that the suppression of electrons from heavy quark (charm
and bottom) decays is comparable to that of light hadrons in central
Au+Au collisions. This implies that heavy quarks may lose a
substantial amount of energy, in contradiction with current
theoretical predictions based on energy loss via induced gluon
radiation for heavy quarks~\cite{armesto,djordjevic}. Recent
calculations including both gluon radiation and collisional energy
loss change the heavy quark energy loss~\cite{mustafa,adil}.
However, the detailed comparison with experimental non-photonic
electron energy loss depends on the relative contributions from
charm and bottom quarks. We have developed an innovative method
which uses the azimuthal correlations between non-photonic electrons
and charged hadrons to estimate the relative $D$ and $B$
contributions to non-photonic electrons~\cite{lin}. Our method
depends on the fact that for the same electron transverse momentum
the near-side e-h angular correlation from $B$ decays is much wider
than that from D decays. In this paper we will present the
preliminary results on the measurement of azimuthal correlations
between high-$p_{T}$ non-photonic electrons and charged hadrons in
p+p collisions at 200 GeV from STAR. We will use comparisons of the
experimental results with PYTHIA simulations to estimate $B$ and $D$
decay contributions to non-photonic electrons as a function of
$p_{T}$ for $p_{T} > 2.5$ GeV/c.

\section{Data analysis and results}
\vspace{-0.3cm}

\begin{figure}[htb]
\begin{center}
                 \includegraphics[width=2.8in]{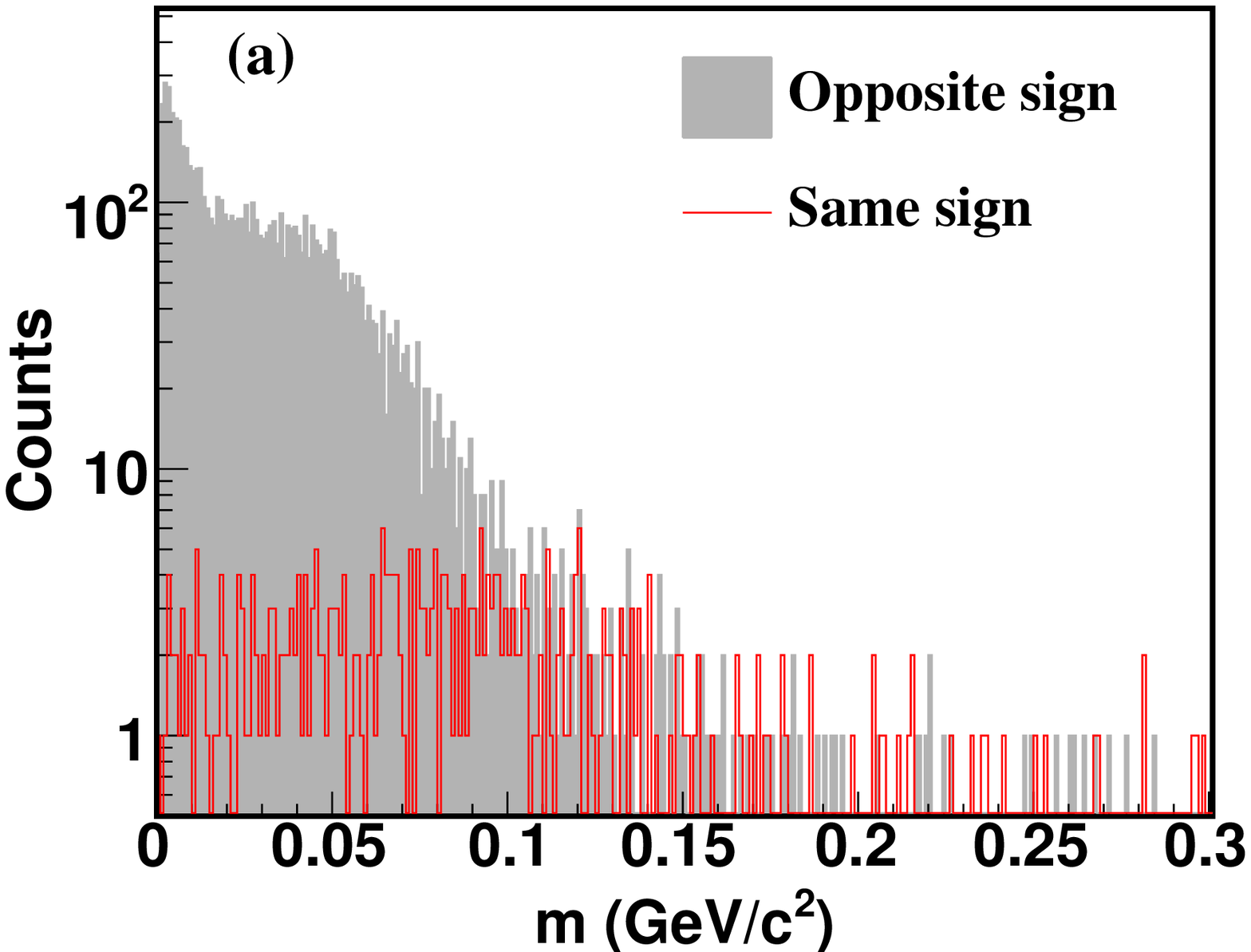}%
                 \includegraphics[width=2.8in]{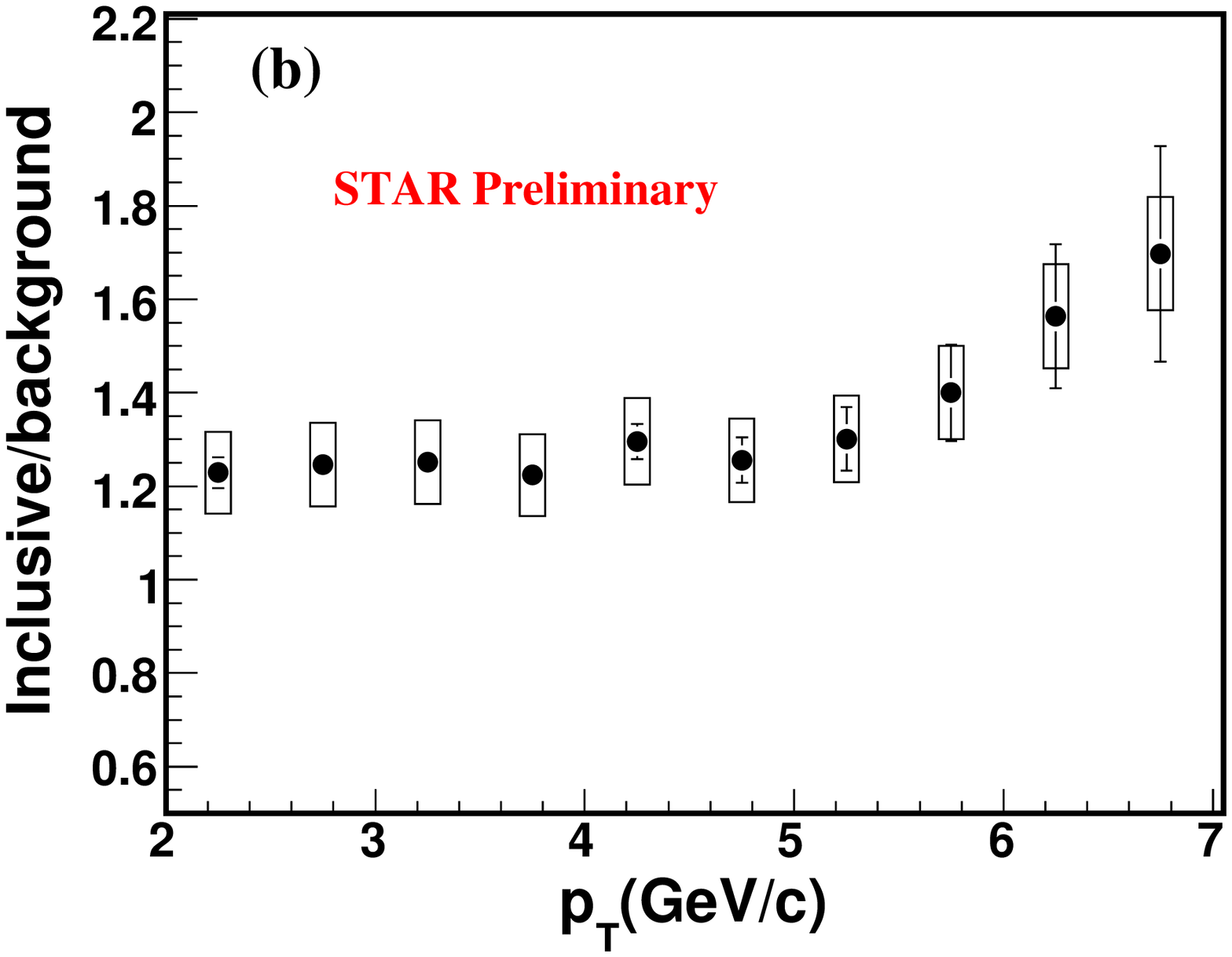}
\vspace*{-.3cm} \caption[]{(a)Invariant mass distributions of
$e^{+}e^{-}$ pairs ($OppSign$, grey filled area) and combinatorial
background ($SameSign$, red solid line) in p+p collisions. (b)The
ratio of inclusive electron to photonic background as a function of
$p_{T}$ in p+p collisions.} \label{fig1} \vspace*{-.3cm}
\end{center}
\end{figure}

The data used in this analysis is p+p events at $\sqrt{s_{NN}} =
200$ GeV recorded by STAR in RUN
\uppercase\expandafter{\romannumeral5}. The main detectors utilized
in this analysis are the STAR Time Projection Chamber
(TPC)~\cite{tpc} and the STAR Barrel Electromagnetic Calorimeter
(BEMC) with the Shower Maximum Detector (SMD)~\cite{emc}. To obtain
sufficient statistics at high-$p_{T}$, we used high tower triggers
corresponding to an energy deposition of at least 2.6 GeV (HT1) and
3.5 GeV (HT2) in a single tower of the BEMC. Around 2.4 million HT1
events and 1.7 million HT2 events were used for the results
presented in this paper.

Electron identification was carried out by combining ionization
energy loss in the TPC with energy deposition in the EMC and shower
profile in the SMD. Detailed descriptions of the electron
reconstruction techniques can be found in Ref.~\cite{star_e,dong}.
The dominant photonic electron background is from photon conversions
and $\pi^{0}$, $\eta$ Dalitz decays, whose electron pairs have small
invariant masses. We combine the electron candidates with tracks
passing a very loose cut on $dE/dx$ around the electron band. The
invariant mass distribution of $e^{+}e^{-}$ pairs ($OppSign$) is
depicted by the grey filled area in panel (a) of Fig.~\ref{fig1}.
The $OppSign$ contains the true photonic background as well as the
combinatorial background, where non-photonic electrons may be
falsely identified as photonic electrons. The combinatorial
background, which is small in p+p collisions, can be estimated by
calculating the invariant mass of same-sign electron pairs
($SameSign$) shown as red solid line in panel (a) of
Fig.~\ref{fig1}. A cut of mass $< 0.1$ GeV$/c^{2}$ rejects about
$70\%$ of photonic background. Panel (b) of Fig.~\ref{fig1} shows
the ratio of inclusive electron to photonic background as a function
of $p_{T}$. The bars (boxes) on the data indicate the size of
statistical (systematic) errors. A significant excess of electrons
with respect to the background can be observed, where the excess
electrons are mostly from heavy quark semi-leptonic decays. STAR has
larger amount of material in RUN
\uppercase\expandafter{\romannumeral5} than RUN
\uppercase\expandafter{\romannumeral3}
(\uppercase\expandafter{\romannumeral4}), leading to inclusive
electron to photonic background ratios from RUN
\uppercase\expandafter{\romannumeral5} systematically lower than
those from RUN \uppercase\expandafter{\romannumeral3}
(\uppercase\expandafter{\romannumeral4})~\cite{star_e}.

\begin{figure}[htb]
\begin{center}
                 \includegraphics[width=2.8in,height=1.8in]{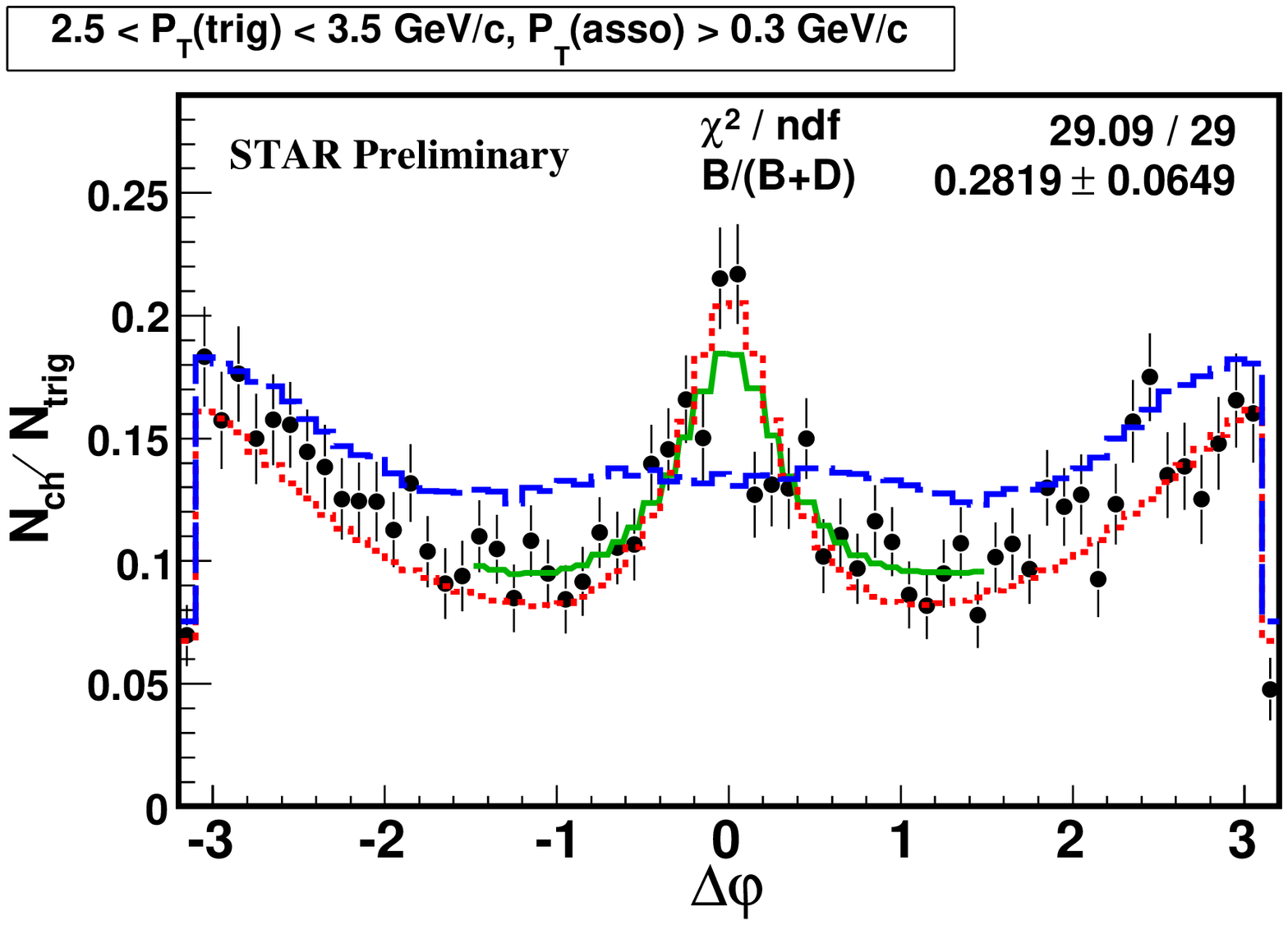}%
                 \includegraphics[width=2.8in,height=1.8in]{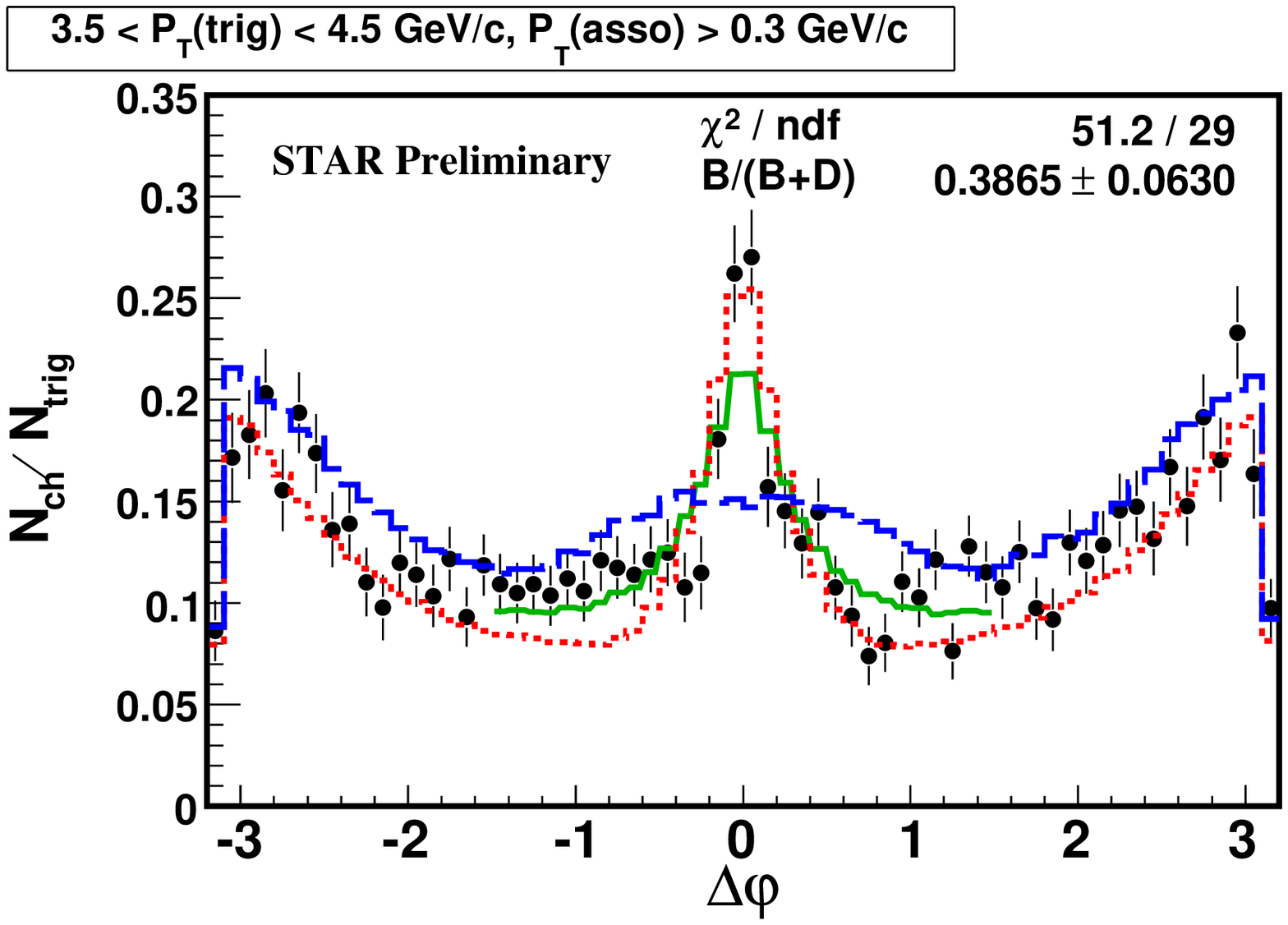}
                 \includegraphics[width=2.8in,height=1.8in]{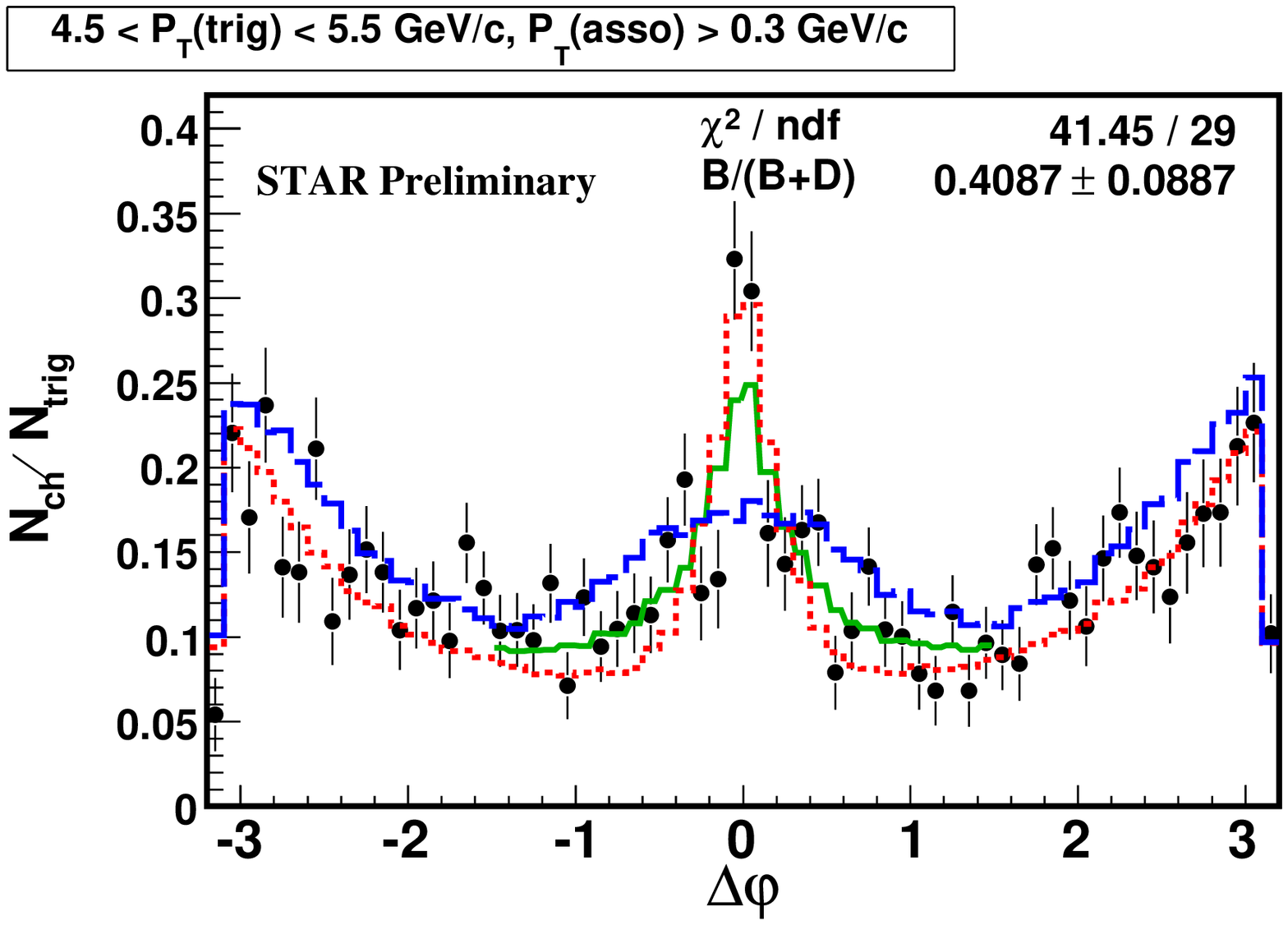}%
                 \includegraphics[width=2.8in,height=1.8in]{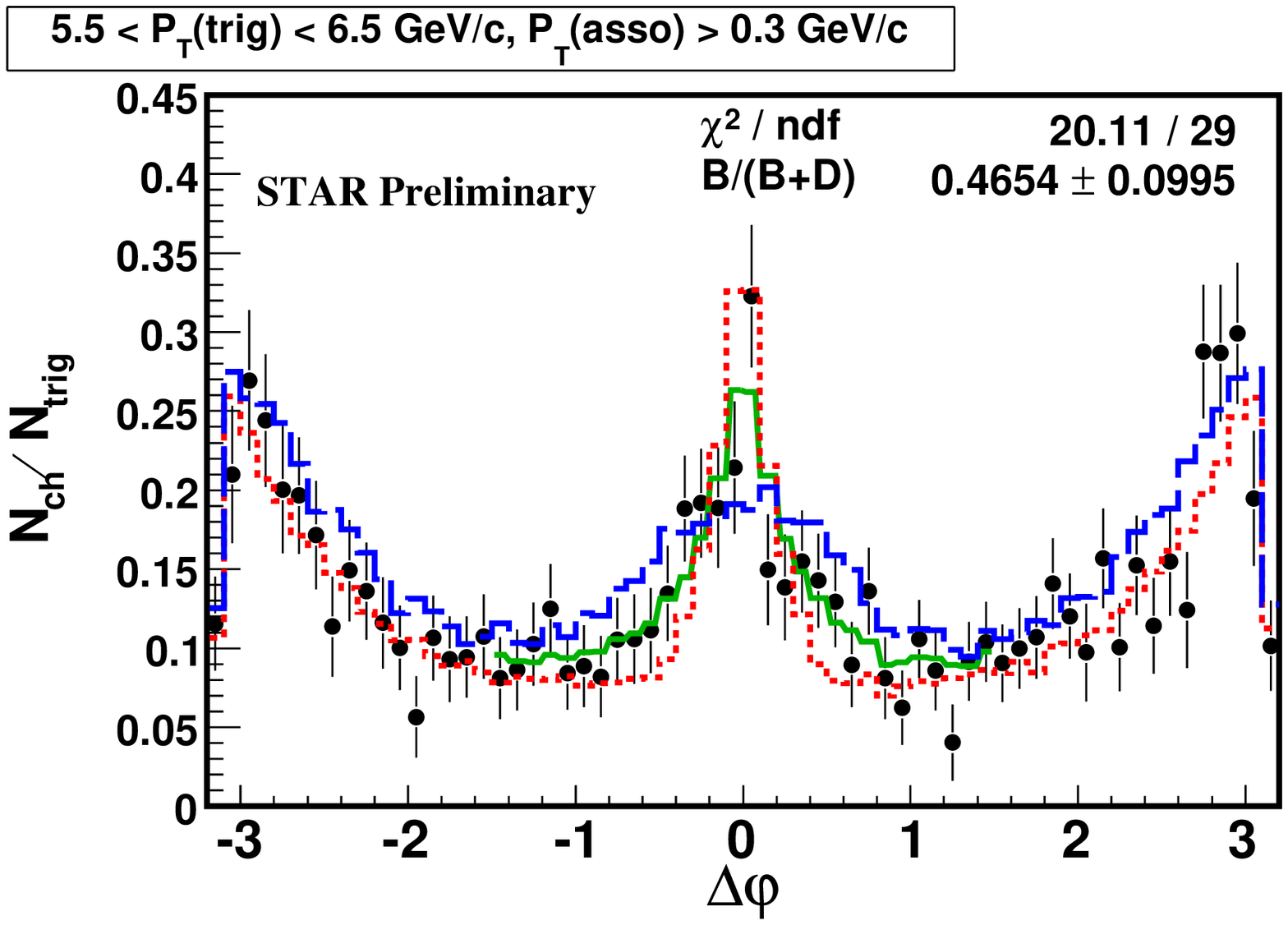}
\vspace*{-.3cm} \caption[]{$\Delta\varphi_{non-pho}$ distributions
and the comparison to PYTHIA simulations for four electron trigger
cuts with associated hadron $p_{T}(assoc)
>$ 0.3 GeV/c. The data are shown as dots, the simulations are depicted by blue dashed lines for $B$ meson
decays and red dotted lines for $D$ meson decays. The green solid
curves are the fits to data points using PYTHIA curves.}
\label{fig2} \vspace*{-.3cm}
\end{center}
\end{figure}

In order to extract the angular correlation between non-photonic
electrons and charged hadrons, we start with the semi-inclusive
electron sample. We remove the $OppSign$ sample after the mass cut
from the inclusive electron sample. The remaining electrons form the
"semi-inclusive" electron sample. The relationship of these samples
is: {\it semi-inc} $=$ {\it inc} - {\it OppSign with the mass cut}
$=$ {\it inc} $-$ ({\it reco-pho} $+$ {\it combinatorics}) $=$ {\it
inc} $-$ ({\it pho} $-$ {\it not-reco-pho} $+$ {\it combinatorics})
$=$ {\it non-pho} $+$ {\it not-reco-pho} $-$ {\it combinatorics}.
Therefore the signal can be obtained by the equation:
$\Delta\varphi_{non-pho} = \Delta\varphi_{semi-inc} -
\Delta\varphi_{not-reco-pho} + \Delta\varphi_{combinatorics}$.
$\Delta\varphi_{not-reco-pho}$ can be calculated using
$\Delta\varphi_{reco-pho}$ by an efficiency correction after
removing the photonic partner of the reconstructed-photonic
electron. The reason to remove the photonic partner is that for the
reconstructed-photonic electron the photonic partner is found while
for not-reconstructed-photonic electron the partner is missing. The
resulting e-h correlations for reconstructed photonic electrons and
not reconstructed photonic electrons cannot be related to each other
by an efficiency correction factor alone. Therefore
$\Delta\varphi_{not-reco-pho}$ can be obtained by the equation:
$\Delta\varphi_{not-reco-pho} = (1/\varepsilon -
1)\Delta\varphi_{reco-pho-no-partner}$, where $\varepsilon$ is the
photonic electron reconstruction efficiency and {\it
reco-pho-no-partner} means reconstructed photonic electrons after
removing the photonic partner. The corresponding efficiency can be
calculated from simulations and is $\sim 70\%$.

Fig.~\ref{fig2} shows the $\Delta\varphi_{non-pho}$ distributions
and the comparison to PYTHIA simulations for four electron trigger
cuts with associated hadron $p_{T}(assoc)
>$ 0.3 GeV/c. The data are shown as dots, the blue dashed curves and the
red dotted curves are from PYTHIA simulations for $B$ decays and for
$D$ decays, respectively. The setup of PYTHIA parameters is
discussed in Ref.~\cite{lin}. We use PYTHIA curves to fit the data
points with the $B$ contribution as a parameter in the fit function:
$\Delta\varphi_{exp} = R\times \Delta\varphi_{B} + (1-R)\times
\Delta\varphi_{D}$, where $R$ is the $B$ contribution, $B/(B+D)$.
The fits are shown as green solid curves in Fig.~\ref{fig2}. The fit
results are consistent within statistical errors when we vary the
fit range in $\Delta\varphi$ from $\pm 1$ to $\pm \pi$. As a
systematic check, we allowed an overall normalization factor in the
fit function to float. It gives a normalization factor close to
unity, and consistent $B/(B+D)$ ratios.

\begin{wrapfigure}{l}{0.45\textwidth}
\vspace{-1.0cm}
\includegraphics[width=3.0in]{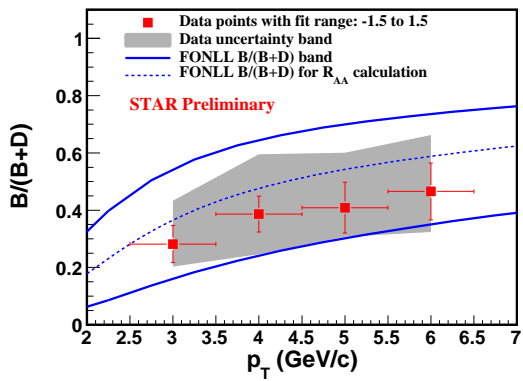}
\caption[]{The $B$ contribution to non-photonic electrons as a
function of electron $p_{T}$.} \label{fig3} \vspace{-0.3cm}
\end{wrapfigure}

The $B$ semi-leptonic decay contribution to non-photonic electrons
as a function of $p_{T}$ is shown in Fig.~\ref{fig3}. The bars show
the size of statistical errors. The grey band indicates the data
uncertainties including statistical errors, and systematic
uncertainties introduced by photonic electron reconstruction
efficiency (dominant), different fit ranges and different fit
functions. The blue solid curves show the range of relative bottom
contribution from recent pQCD calculations (FONLL)~\cite{cacciari}.
The dashed line is the $B/(B+D)$ ratio used for default non-photonic
electron $R_{AA}$ calculation in FONLL. The preliminary STAR data
are consistent with the FONLL calculation.

\section{Conclusions}
\vspace{-0.3cm}

In conclusion, the first measurement of $B$ contribution to
non-photonic electrons has been presented, using the azimuthal
correlations between non-photonic electrons and charged hadrons in
p+p collisions at $\sqrt{s_{NN}}=200$ GeV from STAR. Within the
present statistical and systematic errors, the preliminary data
analysis based on PYTHIA model indicates at $p_{T} \sim 4.0 - 6.0$
GeV/c the measured $B$ contribution to non-photonic electrons is
comparable to $D$ contribution and that the measured $B/(B+D)$
ratios are consistent with the FONLL theoretical calculation.
Together with the observed suppression of non-photonic electrons in
central Au+Au collisions, the measured $B/(B+D)$ ratios imply that
bottom may be suppressed in central Au+Au collisions at RHIC.

\vspace{.5cm} \noindent This work is partly supported by NSFC under
the grant number 10575042, 10610285.


\section*{References}
\vspace{-0.3cm}

\begin{thebibliography}{10}
\bibitem{star_e} Abelev B I {\it et al} (STAR Collaboration) 2006 {\it preprint} nucl-ex/0607012.
\bibitem{phenix_e} Adler S S {\it et al} (PHENIX Collaboration) 2006 {\it Phys. Rev.
Lett.} {\bf 96} 032301.
\bibitem{armesto} Armesto N {\it et al} 2005 {\it Phys. Rev. D}
{\bf 71} 054027.
\bibitem{djordjevic} Djordjevic M {\it et al} 2006 {\it Phys. Lett.
B} {\bf 632} 81.
\bibitem{mustafa} Mustafa M G 2005 {\it Phys. Rev. C} {\bf 72}
014905.
\bibitem{adil} Adil A {\it et al} {\it preprint} nucl-th/0606010.
\bibitem{lin} Lin X Y 2006 {\it preprint} hep-ph/0602067.
\bibitem{tpc} Anderson M {\it et al} 2003 {\it Nucl. Instr. Meth.
A} {\bf 499} 659.
\bibitem{emc} Beddo M {\it et al} 2003 {\it Nucl. Instr. Meth. A}
{\bf 499} 725.
\bibitem{dong} Dong W J 2006 {\it Ph.D. thesis, UCLA}.
\bibitem{cacciari} Cacciari M {\it et al} 2005 {\it Phys. Rev.
Lett.} {\bf 95} 122001; FONLL calculations with CTEQ6M, $m_{c} =
1.5$ GeV/$c^2$, $m_{b} = 5$ GeV/$c^2$, and $\mu_{R,F} = m_{T}$.
\end{thebibliography}
\end{document}